\documentclass[3p,twocolumn]{revtex4}

	\newcommand{\hl}[1]{{#1}}
	\newcommand{\hn}[1]{{#1}}
	\usepackage{amsmath}
	\usepackage{amsfonts}
    \usepackage{amssymb}
	\usepackage{amsfonts}
	\usepackage[ansinew]{}

\newcommand{\be}{\begin{equation}}
\newcommand{\ee}{\end{equation}}
\newcommand{\bea}{\begin{eqnarray}}
\newcommand{\eea}{\end{eqnarray}}

\newcommand\SPD{\mathrel{\stackrel{\makebox[0pt]{\mbox{\normalfont\tiny (3)}}}{\Delta}}}

\newcommand{\R}{\mathcal{R}}

\newcommand{\dd}{{\rm d}}

\usepackage{xcolor}

\newcommand{\SA}[1]{{ #1}}

\makeindex

\begin{document}

\title{
The effect of anisotropic stress and non-adiabatic pressure perturbations on the 
evolution of the comoving curvature perturbation
}

\author{Atsushi Naruko$^{3,5}$}

\author{Antonio Enea Romano$^{1,2,7}$}

\author{Misao Sasaki$^{4,5,6}$}

\author{Sergio Andr\'es Vallejo-Pe\~na$^{1}$}

\affiliation{$^{1}$Instituto de Fisica, Universidad de Antioquia, A.A.1226, Medellin, Colombia}

\affiliation{$^{2}$Theoretical Physics Department, CERN, CH-1211 Geneva 23, Switzerland} 

\affiliation{$^{3}$Frontier Research Institute for Interdisciplinary Sciences \& Department of Physics, \\
Tohoku University, Sendai 980-8578, Japan}

\affiliation{$^{4}$Kavli Institute for the Physics and Mathematics
of the Universe (WPI), University of Tokyo, Kashiwa 277-8583, Japan}

\affiliation{$^{5}$Center for Gravitational Physics, Yukawa Institute for Theoretical
Physics, Kyoto University, Kyoto 606-8502, Japan}

\affiliation{$^{6}$Leung Center for Cosmology and Particle Astrophysics, National Taiwan University, Taipei 10617, Taiwan}

\affiliation{$^{7}$Department of Physics \& Astronomy, Bishop’s University
2600 College Street, Sherbrooke, Qu´ebec, Canada J1M 1Z7
}

\begin{abstract}
We derive the 
equation  for the evolution of the curvature perturbation on the comoving time slice, $\R_c$,  in the presence of anisotropic and non-adiabatic terms in the energy-momentum tensor of matter fields.  
The equation is obtained by manipulating the perturbed Einstein's equations in the comoving time slice. 
It could be used to study the evolution of the comoving curvature perturbations for systems with an anisotropic energy-momentum tensor, such as in the presence of vector fields, \hn{in the presence of entropy, such as in a multi-field system, or in modified gravity theories}. 
As a simple application, after checking that the comoving time slice for a multi-field system does not coincide with the uniform field time slice \hl{in general}, we \hl{use  the equation in } the case of two minimally coupled scalar fields and derive a closed set of equations for the curvature and entropy perturbations on the comoving time slice.
\end{abstract}


\thispagestyle{myheadings}{TU-1059, YITP-17-116,IPMU18-0075}

\maketitle
\section{Introduction}
The theory of cosmological perturbations is very useful to study the early stages of the Universe, especially during inflation, that is, an exponential expansion phase of the Universe which the standard cosmological model hypothesizes to explain observations such as anisotropies in the cosmic microwave background radiation (CMB). 
One quantity which is particularly important in this context is the curvature perturbation on the comoving slice, $\R_c$. 
In single field slow-roll inflation models this quantity is conserved on super-horizon scales\cite{Romano:2015vxz,Wands:2000dp}, which has important  implications on the relation between primordial perturbations and late-time observables such as the CMB anisotropies. 
For a globally adiabatic system in a single field model this quantity may not be conserved \cite{Romano:2016gop}. Other possible causes of the super-horizon 
evolution of perturbations could be anisotropic stress or \hl{non-adiabatic pressure} components of the energy-momentum tensor.

 In this short note we derive the equations for 
the curvature perturbation on the comoving time slice, $\R_c$, including 
two terms, namely anisotropic stress and \hl{non-adiabatic pressure} terms, showing that they act, as expected, as source terms which can be relevant even on super-horizon scales. Our approach is quite generic and can be applied
to any system which can be described by an energy-momentum tensor of the form we use, not only to a multi-scalar system but also a system with vector fields. 
The derivation is based on manipulating the Einstein equations in order to obtain an equation involving only $\R_c$, 
the anisotropic stress and \hl{non-adiabatic pressure} terms and background quantities. 
The equation can be used to study phenomenologically the effect of anisotropic stress tensor and non-adiabaticity without assuming any specific model.
One useful application could be to study models which violate the non-Gaussianity consistency relation \cite{Maldacena:2002vr} that was derived in fact based on the assumption of the conservation of 
the comoving curvatue perturbation on superhorizon scales.

\section{Evolution of comoving curvature perturbations}

The Einstein equations in a spatially flat Friedmann-Lema\^itre-Robertson-Walker (FLRW) background are
\begin{align}
 3 \mathcal{H}^2 &= a^2 \, \rho \,, \\
 2 (\mathcal{H}'-\mathcal{H}^2) &= - a^2 \, (\rho+P)  \,.  \label{faeq}
\end{align}
Here a prime denotes a derivative with respect to the conformal time $\eta$ and 
 $\mathcal{H}$ stands for the conformal Hubble parameter defined by $\mathcal{H} = a'/a$.
$\rho$ and $P$ represent the background energy density and pressure of the matter field respectively.
We use the units in which $8\pi G=c=1$.

Scalar perturbations on a spatially flat FLRW metric can be written as 
\begin{align}
ds^2 = a^2 \, &\Bigl[ -(1+2A) \dd\eta^2+2\partial_iB \dd x^i\dd\eta+ \nonumber \\ 
& \quad + \bigl[ (1+2\mathcal{R})\delta_{ij} + 2\partial_i\partial_jE \bigr] \dd x^i \dd x^j \Bigr] \,, 
\label{pmetric}
\end{align}
 where the Latin indices run from $1$ to $3$.
The corresponding energy-momentum tensor takes the form :
\begin{align}
T^0{}_0 &= - (\rho+\delta\rho) \,, \quad T^0{}_i = \frac{\rho+P}{a}u_i \,, \nonumber \\ T^i{}_j &= (P+\delta P)\delta^i{}_j+\Pi^i{}_j \,,
\label{psem}
\end{align}
where  
\begin{align}
u_i &= a \, \partial_i(v+B) \,, \label{pcv} \\
\Pi^i{}_j&= \delta^{ik}
\partial_{k}\partial_{j}\Pi
-\frac{1}{3} \SPD \Pi \delta^{i}{}_{j} \,, \quad \Pi^i{}_i = 0 \,. \label{anstress}
\end{align}
In the above equations $\Pi^{i}{}_{j}$ is the \hl{anisotropic stress}, $v$ is the velocity potential, $\Pi$ is 
the anisotropy potential and we have defined 
$\SPD \equiv \delta^{ij}\partial_i\partial_j$.
\hl{Note that any energy-momentum tensor can be decomposed according to Eq.~(\ref{psem}), making all the results derived from it completely general and applicable to any system with a well defined energy-momentum tensor, including muti-fields, and modified gravity theories.}

The curvature perturbation on the comoving slice $\mathcal{R}_c$ is a gauge-invariant quantity defined as the curvature perturbation $\mathcal{R}$ evaluated on the hypersurfaces in which $v+B$ vanishes. 
The spatial Fourier expansion of the linearly perturbed Einstein equations on the comoving slice \cite{Malik2009} takes the form :
\begin{align}
2 k^2 (\mathcal{R}_{c} - \mathcal{H}\sigma_{c}) 
&= a^2 \delta \rho_{c}  \, , \label{eqzz}\\ 
\mathcal{R}_{c}' - \mathcal{H} A_{c} &= 0 \, , \label{eqzi} \\
 2 (\mathcal{H}' -\mathcal{H}^2) A_{c}
 &= a^2 \Bigl[ \delta P_{c} - (2 k^2/3) \Pi_{c} \Bigr] \, , \label{eqii} \\
\sigma_{c}'+2\mathcal{H}\sigma_{c}-A_{c}-\mathcal{R}_{c} &= a^2 \Pi_{c} \, , \label{eqij}
\end{align}
where $\sigma=E'-B$ is the scalar shear.
In general we can decompose the pressure perturbation as
\begin{equation}
\delta P_c=c_s^2(\eta) \, \delta \rho_c+\Gamma_c \,, \label{dP}
\end{equation}
where we can interpret $c_s$ and $\Gamma_c$ as the adiabatic sound speed and
the non-adiabatic part of the pressure perturbation respectively.
For a minimally coupled single scalar field model $c_s=1$ and $\Gamma_c$ is zero, 
but in general one would expect that $\Gamma_c$ could be non-vanishing.
Our goal is to derive  an equation for $\mathcal{R}_{c}$ in the presence of 
both anisotropic stress $\Pi^i{}_j$ and non-adiabatic pressure $\Gamma_c$ perturbations.

First we use Eq.~(\ref{eqzi}) to express $A_c$ in terms of $\mathcal{R}_{c}$, 
\begin{equation}
A_c= \frac{\mathcal{R}_{c}'}{\mathcal{H}} \label{eqA} \, .
\end{equation}
We substitute this $A_c$ and $\delta P_c$ given in Eq.~(\ref{dP}) into 
Eq.~(\ref{eqii}), and solve it for $\delta\rho_c$ :
\begin{equation}
\delta \rho_{c}
=\frac{1}{c_s^2} \left( \frac{2}{3} k^2 \Pi_c - \Gamma_c 
- \frac{\rho + P}{\mathcal{H}} \mathcal{R}_c' \right) \, . \label{drho}
\end{equation}
We then insert this  into Eq.~(\ref{eqzz})  to get an expression for $\sigma_{c}$ : 
\begin{align}
\sigma_{c}&= \frac{1}{\mathcal{H}} \left[ \mathcal{R}_c
- \frac{a^2}{2 k^2 c_s^2} \left( \frac{2}{3} k^2 \Pi_c
- \Gamma_c - \frac{\rho + P}{\mathcal{H}} \mathcal{R}_c' \right) \right] \,. \label{sigma}
\end{align}
Finally we substitute $A_c$ and $\sigma_c$ given by Eqs.~(\ref{eqA}) and (\ref{sigma}) respectively, into Eq.~(\ref{eqij}) to obtain 
\begin{equation}
\mathcal{R}_c''+ 2 \frac{z'}{z} \mathcal{R}_c'
- c_s^2\SPD\mathcal{R}_c + \frac{\mathcal{H}}{\rho +P} Y_c =0 \, , \label{deccp}
\end{equation}
where we have defined
\begin{align}
 z^2 &\equiv \frac{a^4 (\rho+P)}{c_s^2\mathcal{H}^2} \,,  \\ 
Y_c &\equiv  \left[\log \left(\frac{a^4}{\mathcal{H} c_s{}^2}\right)\right]'
\left(\frac{2}{3} \SPD \Pi_c +\Gamma_c \right) \nonumber \\ 
& \qquad {} + 2 \mathcal{H} c_s^2 \SPD \Pi_c + \frac{2}{3} \SPD \Pi_c' + \Gamma_c' \,. \label{zetaeq}
\end{align}
This is the main result of this paper.
As expected, for adiabatic ($\Gamma_c=0$) and isotropic perturbations ($\Pi_c=0$)
 the above equation takes the well-known form :
\begin{equation}
\mathcal{R}_c''+ 2 \frac{z'}{z}\mathcal{R}_c'
- c_s^2\SPD\mathcal{R}_c=0 \,. 
\end{equation}

 \hl{ Since \hn{eq.\eqref{deccp}} has been derived assuming the most general scalar perturbations of the energy-momentum tensor, it can be applied to any system which satisfies the Einstein's equations. 
While we have explicitly used the Einstein's equation, it is possible to rewrite the gravitational field equation in this form even in \hn{modified gravity theories} by identifying the deviation from the Einstein's tensor as \hn{an} effective energy momentum tensor, $G_{\mu\nu}=T^\mathrm{matter}_{\mu\nu} + T^\mathrm{eff}_{\mu\nu}$. \hn{As a consequence} eq.\eqref{deccp} can \hn{also be applied to modified gravity theories, but in this case 
 the effective anisotropic stress and non-adiabatic pressure terms can be functions of  metric perturbations,
 and further manipulation is required to obtain an equation only involving $\R_c$ and matter fields .}}

\section{Curvature perturbation for scalar fields }
Given the generality of the form of the energy momentum tensor used in the derivation of Eq.~(\ref{deccp}) it can be applied to a wide class of physical scenarios, including multi-field systems. Let us consider the case of two minimally coupled  scalar fields  with  Lagrangian 
\begin{align}
L = - \sum^2_{n = 1} X_n - 2 V(\Phi_1,\Phi_2) \, ,
\end{align}
where $X_n=g^{\mu\nu}\partial_\mu\Phi_n\partial_\nu\Phi_n$ and  $\Phi_n(x^{\mu})= \phi_n(\eta) + \delta \phi_n(x^{\mu})$.
The perturbed energy-momentum tensor, without gauge fixing, is given by 
\begin{align}
\delta \rho &= \frac{\phi_1' \delta\phi_1' + \phi_2' \delta\phi_2' - A (\phi_1'{}^2+\phi_2'{}^2)}{a^2} + V_1  \delta \phi_1 + V_2 \delta \phi_2 \nonumber \, ,  \\
\delta P &= \frac{\phi_1' \delta\phi_1' + \phi_2' \delta\phi_2' - A (\phi_1'{}^2+\phi_2'{}^2)}{a^2} - V_1 \delta \phi_1 - V_2 \delta \phi_2  \, ,  \nonumber  \\
\Pi&=0 \, \quad , \quad  \delta T^{0}{}_{i} = \partial_i \left( -\frac{\phi_1' \delta \phi_1+\phi_2' \delta \phi_2}{a^2} \right) \, , \label{pq}
\end{align}
where we denote the partial derivatives as \SA{$V_n = (\partial V/\partial \Phi_n) (\phi_1,\phi_2)$}. 

The field perturbations transform under an infinitesimal time translation $\eta \to \eta+\delta \eta$ 
\begin{align}
\widetilde{\delta \phi_1} &= \delta \phi_1 - \phi_1' \delta \eta \quad , \quad \widetilde{\delta \phi_2} = \delta \phi_2 - \phi_2'\delta \eta \, . \label{gtdeltapsi}
\end{align}
The time translation $\delta \eta_c$ necessary to define the comoving slices can be found by imposing the condition $(\delta T^{0}{}_{i})_c \propto \phi_1'\delta\widetilde{\phi_1}+\phi_2'\delta\widetilde{\phi_2} = 0$, giving
\begin{align}
\delta \eta_c &= \frac{\phi_1' \delta \phi_1+\phi_2'\delta \phi_2}{\phi_1'{}^2+\phi_2'{}^2} \, . \label{cgt}
\end{align}
Note that the uniform field time slicing, also known as the unitary time slicing, in general does not coincide with the comoving time slicing except for the case of a single field or the case with $\delta \phi_2 \propto \delta \phi_1$. 
The comoving curvature perturbation for the two scalar fields system is given by
\begin{align}
 \R_c = \R - \mathcal{H} \delta \eta_c = \R - \mathcal{H}\frac{\phi_1' \delta \phi_1+\phi_2'\delta \phi_2}{\phi_1'{}^2+\phi_2'{}^2} \,\label{Rcphi}.
\end{align}
The gauge invariant field perturbations in the comoving slices now can be defined 
\begin{align}
U_{1} &= \delta \phi_1 - \phi_1' \frac{\phi_1' \delta \phi_1+\phi_2'\delta \phi_2}{\phi_1'{}^2+\phi_2'{}^2} \, , \label{U1}\\  
U_{2} &= \delta \phi_2 - \phi_2' \frac{\phi_1' \delta \phi_1+\phi_2'\delta \phi_2}{\phi_1'{}^2+\phi_2'{}^2} \, , \label{U2}
\end{align}
and similarly for the  pressure and energy perturbations in the comoving slices we get
\begin{align}
\delta \rho_c &=\frac{\phi_1' U_1' + \phi_2' U_2' - A_c (\phi_1'{}^2+\phi_2'{}^2)}{a^2} + V_1 U_{1} +V_2 U_{2} \label{drhoU} \,, \\
\delta P_c&= \frac{\phi_1' U_1' + \phi_2' U_2' - A_c (\phi_1'{}^2+\phi_2'{}^2)}{a^2} - V_1 U_{1} -V_2 U_{2} \, . \label{dPU}
\end{align}
Note that all the above quantities are gauge invariant by construction.

Combining Eqs.~(\ref{U1}), (\ref{U2}) and the background field equations of motion we find
\begin{align}
\frac{\phi_1' U_1' + \phi_2' U_2'}{a^2}= V_1 U_1 + V_2 U_2 = -  \frac{\phi_1'{}^2+\phi_2'{}^2}{4a^2}\Theta \, , \label{UpU} 
\end{align}
where we have defined the function $\Theta$ according to
\begin{align}
\Theta &= \left[ \frac{\partial}{\partial \eta} \left( \frac{\phi_1'{}^2-\phi_2'{}^2}{\phi_1'{}^2+\phi_2'{}^2} \right) \right] \left(\frac{\delta\phi_1}{\phi_1'} - \frac{\delta\phi_2}{\phi_2'} \right) \, . \label{theta}
\end{align}
Assuming a classical field trajectory parameterized as $\phi_2 = \phi_2(\phi_1)$ we can write $\Theta$  in this form
\begin{eqnarray}
\Theta= - 4 \frac{\dd^2 \phi_2}{\dd \phi_1^2} \left[ \left( \frac{\dd \phi_2}{\dd \phi_1} \right)^2 + 1 \right]^{-2}
\left( \frac{\dd \phi_2}{\dd \phi_1} \delta \phi_1 
- \delta \phi_2 \right)
\end{eqnarray}
From the above expression we can see that in order for $\Theta$ to be different from zero the trajectory has to have non vanishing first and second derivatives, i.e. there must be some turn in the field space.

After replacing Eqs.~(\ref{eqzi}) \SA{and (\ref{UpU}) into Eqs.~(\ref{drhoU}) and (\ref{dPU})} we get  
\begin{align}
\delta \rho_c &= - \frac{\phi_1'{}^2+\phi_2'{}^2}{a^2\mathcal{H}}\mathcal{R}_c' -\frac{\phi_1'{}^2+\phi_2'{}^2}{2a^2}\Theta \, , \label{deltarhoc} \\
\delta P_c &= - \frac{\phi_1'{}^2+\phi_2'{}^2}{a^2\mathcal{H}}\mathcal{R}_c' \,. \label{deltaPc} 
\end{align}
It follows from Eqs.~(\ref{deltarhoc}) and (\ref{deltaPc}) that 
\begin{equation}
\delta P _c = \delta \rho_c+ \frac{\phi_1'{}^2+\phi_2'{}^2}{2a^2} \Theta \, , 
\end{equation}
and comparing this  with Eq.~(\ref{dP}) we obtain the sound speed and the entropy perturbations
\begin{equation}
c_s^2(\eta)=1 \,, \qquad 
\Gamma_c = \frac{\phi_1'{}^2+\phi_2'{}^2}{2a^2}\Theta\, . 
\end{equation}
From these relations we can find a closed system of equations to describe the evolution of $\R_c$  and $\Gamma_c$ 
\begin{align}
&\mathcal{R}_c''+ 2 \frac{z'}{z}\mathcal{R}_c'
-\SPD\mathcal{R}_c  =- \frac{a^2\mathcal{H}}{(\phi_1'{}^2+\phi_2'{}^2)} Y_c \, , \label{RcG} \\  
&\Gamma_c''+\SA{a_c} \Gamma_c' - \SPD \Gamma_c + \SA{b_c}\Gamma_c  =\SA{d_c} \R'_c \,, \label{GammaEq}
\end{align}
where
\begin{align}
z^2 &= \frac{a^2 (\phi_1'{}^2+\phi_2'{}^2)}{\mathcal{H}^2} \, , \\ 
Y_c&= \left[\log \left(\frac{a^4}{\mathcal{H}}\right)\right]'
\Gamma_c  + \Gamma_c' \, ,
\end{align}
and the coefficients $\{a_c,b_c,d_c\}$ in Eq.(\ref{GammaEq}) are given in the appendix.

Eq.~(\ref{RcG}) is in agreement with \cite{Bellido:1995}, confirming that  Eq.~(\ref{deccp}) is general and can also be applied to multi-field systems once the entropy has been appropriately defined.
Note that the approach we adopted to derive Eq.~(\ref{GammaEq}) does not require the decomposition of field perturbations in components parallel and perpendicular to the classical field trajectory as done in \cite{Gordon:2000hv}, but is just based on the fundamental  definition of non-adiabatic pressure given in Eq.~(\ref{dP}). 
 For multi-field systems  the presence of entropy perturbations is a consequence of the fact that the comoving slices and the uniform field slicies do not coincide, contrary to the single field case. In general in order to use Eq.~(\ref{deccp}) it is first necessary to compute the energy momentum tensor in the comoving slices with a procedure similar to the one shown above for two fields.

%

\section{Conclusions}
We have derived a general equation for the evolution of the curvature perturbation on the comoving time slicing taking into account the effect of 
anisotropic and non-adiabatic stress perturbations. 
The equation can be applied also to multi-field systems. 
This approach does not require the decomposition of field perturbations in components parallel and perpendicular to the classical field trajectory, but is based just on the fundamental definition of non-adiabatic pressure. 

As an application we have derived 
a closed system of equations for the curvature and entropy perturbations in the comoving time slice for  two minimally coupled scalar fields. The equations are consistent with the ones obtained using the flat time slice \cite{Gordon:2000hv}.
In future it will be interesting to apply the equation to more generic systems where both \hl{anisotropic stress and non-adiabatic pressure perturbations are present,  such as multi-fields vector models, modified gravity theories, or a combination of the two.}


\acknowledgments
A.N. would like to thank the Yukawa Institute for Theoretical Physics at Kyoto University, where discussions during the YITP symposium YKIS2018a ``General Relativity -- The Next Generation --" were useful to complete this work.
The work of A.N. is supported in part by a JST grant ``Establishing a Consortium for the Development of Human Resources in Science and Technology'' and also supported by the JSPS Grant-in-Aid for Scientific Research No.16H01092.
The work of M.S. is supported in part by MEXT KAKENHI Nos.~15H05888 and 15K21733.
This work was supported by the UDEA Dedicacion exclusiva and
Sostenibilidad programs and the CODI projects 2015-4044 and 2016-10945.

\appendix
\section{The coefficients of the equation for $\Gamma_c$}
\begin{align}
a_c &= 2 \mathcal{H} - 2 a^2 \frac{\phi_1' V_1 + \phi_2' V_2 }{\phi_1'{}^2 + \phi_2'{}^2}
\notag\\
& \quad 
+ 2 \frac{\phi_1'\phi_2'(V_{11} - V_{22})
- (\phi_1'{}^2 - \phi_2'{}^2)V_{12}  }{\phi_1' V_2 - \phi_2' V_1} \,, \\
b_c &= - 6 a^2 \mathcal{H} \frac{\phi_1' V_1 + \phi_2' V_2 }{\phi_1'{}^2 + \phi_2'{}^2}
\notag\\
& \quad 
+ 2 a^2 \frac{2 (\phi_1' V_1 - \phi_2' V_2) V_{12}
- (\phi_1' V_2 + \phi_2' V_1)(V_{11} - V_{22})}{\phi_1'V_2 - \phi_2'V_1}
\notag\\
 & \quad 
+ 2 \left[ \frac{(\phi_1'{}^2 - \phi_2'{}^2) V_{12}
- \phi_1' \phi_2' (V_{11} - V_{22})}{\phi_1'V_2 - \phi_2'V_1} \right]^2
\notag\\
 & \quad 
+ \phi_1' \phi_2' \frac{\phi_1'V_{111}- \phi_2' V_{222}}{\phi_1'V_2 - \phi_2'V_1}  
\notag\\
 & \quad 
- \frac{\phi_1'(\phi_1'{}^2- 2 \phi_2'{}^2)V_{112}
+ \phi_2' (2 \phi_1'{}^2- \phi_2'{}^2)V_{122}}{\phi_1'V_2 - \phi_2'V_1}  
\,, \\
d_c &= \frac{4 a^2 \left(\phi _1' V_2 
-\phi _2' V_1\right){}^2}{\mathcal{H} \left(\phi_1'{}^2+\phi_2'{}^2\right)}  \, . 
\end{align}

\section{The equation for the entropy perturbations in terms of $\gamma \equiv \phi_2' \delta \phi_1 - \phi_1' \delta \phi_2$}
It is convenient to introduce a quantity 
\begin{equation}
\gamma \equiv \phi_2' \delta \phi_1 - \phi_1' \delta \phi_2 \,.
\end{equation}
In terms of this quantity we get
\begin{eqnarray}
U_1&=&\frac{\phi_2'}{\phi_1'{}^2+\phi_2'{}^2} \gamma \quad , \quad  U_2 = - \frac{\phi_1'}{\phi_1'{}^2+\phi_2'{}^2} \gamma \, . \label{U2g} \\ 
\delta \rho_c  
&=& - \frac{\phi_1'{}^2+\phi_2'{}^2}{a^2\mathcal{H}}\mathcal{R}_c'
 - 2 \frac{\phi_1' V_2 - \phi_2' V_1}{\phi_1'{}^2 + \phi_2'{}^2} \gamma
\, , \\ 
\Theta &=& \frac{4a^2(\phi_1' V_2 - \phi_2' V_1)}{(\phi_1'{}^2+\phi_2'{}^2){}^2} \gamma \, , \\ \Gamma_c &=& 2 \frac{\phi_1' V_2 - \phi_2' V_1}{\phi_1'{}^2 + \phi_2'{}^2} \gamma \, .
\end{eqnarray}
Finally we can write the equation of motion for $\gamma_c$
\begin{equation}
\gamma'' + \alpha_c \gamma' -  \SPD \gamma + \beta_c \gamma =\delta_c \R'_c \,,
\end{equation}
where
\bea
\alpha_c &=& 6 \mathcal{H} + 2 a^2 \frac{\phi_1' V_1 
 + \phi_2' V_2}{\phi_1'{}^2 + \phi_2'{}^2} \, , \\
 \beta_c &=& 10 \mathcal{H}^2 
+ 10 a^2 \mathcal{H} \frac{\phi_1'V_1 + \phi_2' V_2}{ (\phi_1'{}^2+\phi_2'{}^2)} +a^2 (V_{11}+V_{22})  \nonumber \\
 && \quad
- (\phi_1'{}^2 + \phi_2'{}^2)
 + \frac{2 a^4}{{(\phi_1'{}^2+\phi_2'{}^2){}^2}} \Big[4V_1V_2\phi_1'\phi_2'+ \nonumber \\ && \quad +(\phi_1'{}^2-\phi_2'{}^2) (V_1^2 - V_2^2)\Big] \, , \\
 \delta_c &=&  2 a^2  \frac{\phi_1' V_2 - \phi_2' V_1}{\mathcal{H}} \, .
\eea

\bibliographystyle{h-physrev4}
\bibliography{mybib} 
\end{document}